**High-Resolution Magnetic Particle Imaging System Matrix Recovery Using a Vision Transformer with Residual Feature Network**

Abuobaida M.khair[a,b], Wenjing Jiang[a], Yousuf Babiker M. Osman[c], Wenjun Xia[d,*] and Xiaopeng Ma[a,1]

[a] School of Control Science and Engineering, Shandong University, Jinan 250061, China.
[b] Department of Medical Physics, Al-Neelain University, Khartoum, Sudan
[c] Paul C. Lauterbur Research Center for Biomedical Imaging, Shenzhen Institute of Advanced Technology, Chinese Academy of Sciences, Shenzhen 518055, Guandong, China.
[d] Department of Intensive Care Unit, Shandong Provincial Hospital Affiliated to Shandong First Medical University, Jinan 250012, China.

## 1. Introduction

Magnetic particle imaging (MPI) is a novel imaging technology that employs magnetic nanoparticles (MNPs) and nonlinear re-magnetization to detect their local concentration, which indicates that MPI does not provide morphological information [1], [2]. However, MPI has advantages that make it a useful technique for clinical imaging. MPI is expected to be implemented in various fields, such as dynamic imaging in coronary artery diseases and blood vessel visualization; additionally, potential future uses include cancer detection [3], [4].

The image produced by a signal is generated by magnetic nanoparticles in MPI in response to dynamic magnetic fields brought by a sinusoidal drive field. This signal makes the spatial distribution of nanoparticles recognizable, allowing for high-resolution and quantitative imaging. For spatial encoding, the selection field in MPI generates dynamic and saturation areas, enabling particles to respond differently depending on where they are. During the MPI image reconstruction process, the signal is mapped to the spatial domain [5]. Two primary concepts were established to conduct this mapping: (i) utilizing a system matrix (SM) to solve a set of linear equations, and (ii) an approach that directly associates the time signal with a grid termed x-space. Reconstruction can be employed in various ways, depending on the field sequence, scanning device, and tracer material utilized [6]. By measuring the particle signal at predetermined spatial coordinates that will subsequently represent the voxels of the reconstructed images, it is possible to acquire the SM most accurately [7]. The steepness of the gradient field influences the spatial resolution of the resulting image, the slope of the magnetization curve, and the signal-to-noise ratio (SNR) of the measurement data [8]. The SM is exposed to various changes in the overall process of image reconstruction, such as changing MNP size and the magnetic field variation. Therefore, recovery is required to tolerate these irregularities [9]. In this work, the SM will be recovered by implementing a deep learning model consisting of a vision transformer (ViT) combined with a residual feature network and based on the fine features extracted from each point on the SM from each layer, namely Vision Transformer–Residual Feature Network (VRF-Net). We employed two levels of downsampling of the frequency components in each row to degrade the system matrix. We create residual feature blocks that increase feature learning to deal with insufficient feature representation due to downsampling.

Moreover, the resolution of MPI systems is limited by a trade-off with the receive coil sensitivity, as higher resolution may diminish signals. Also, increased noise levels due to decreased sensitivity at greater distances from the coil. To address this, we present a simulation-based dataset that considers varying coil sensitivity profiles. This strategy allows the model to learn the spatial patterns and the subtle fluctuations of coil sensitivity [10][11].

In designing VRF-Net, careful consideration is given to the frequency composition of the system matrix and the degradation processes that affect its recovery. The system matrix in MPI captures a complex spectrum of spatial frequency components that arise from the nonlinear magnetic response of nanoparticles. These frequency components encode vital structural information, and their accurate recovery is essential for producing high-fidelity images. However, reconstructing both global patterns and fine-scale features from degraded or undersampled system matrices remains a central challenge[12].

To address this, VRF-Net integrates a vision transformer module with a residual feature learning pathway. The vision transformer serves as a global reasoning mechanism. Unlike traditional convolutional models, which operate primarily within local neighborhoods, transformers are capable of modeling long-range dependencies across the entire spatial domain. Through attention-based mechanisms, the transformer adaptively learns relationships between spatially distant elements in the system matrix, allowing it to infer structural coherence and mitigate the effects of signal loss or distortion across the field of view. This ability to capture non-local interactions is particularly important in MPI, where particle responses are influenced by large-scale field variations and harmonic content[13].

While the transformer excels at global feature modeling, it is less suited to preserving the subtle local patterns that often define fine image detail. To address this, the model incorporates a residual feature enhancement pathway composed of lightweight convolutional layers. This component acts as a complementary local processor, focusing on refining edge information, small-scale variations, and high-frequency structures. Using residual connections, it reinforces the preservation of input features while enabling deeper learning without vanishing gradients. The fusion of global transformer-based attention with residual local refinement allows the model to maintain a robust understanding of both coarse and fine structures within the degraded system matrix.

To make the model resilient to practical resolution limitations, we simulate degraded input data by applying a dual downsampling process. This is motivated by the need to train the model on inputs that reflect the lower-resolution, noisy, or incomplete system matrices commonly encountered in real MPI acquisitions. Downsampling in this context is not only a data reduction technique; it functions as a deliberate signal degradation strategy that removes selected high-frequency components from the frequency domain. Specifically, the original system matrix is downsampled twice: first to simulate an initial loss of resolution, and then again to mimic further degradation due to spatial encoding variability or coil sensitivity fluctuations [14].

The mechanism involves applying pooling operations along the frequency axis of each system matrix row, which reduces dimensionality while preserving dominant features in a compressed form. The first downsampling stage acts as a controlled reduction, ensuring that the dominant spectral patterns are retained. The second stage introduces additional loss, simulating real-world degradation scenarios and encouraging the model to learn how to reconstruct high-resolution outputs even from heavily simplified inputs. This dual-stage degradation ensures the model is trained under realistic constraints, promoting its ability to generalize to various acquisition conditions. However, it also imposes a reconstruction challenge, recovering fine details from information that may have been partially or fully suppressed. VRF-Net is a hybrid architecture specifically designed to confront this challenge through its joint use of global attention and local residual enhancement [15]. Transformers have recently been introduced for MPI system matrix recovery[16], [17] showing promising results in capturing global dependencies. However, their use has so far been limited to standalone architectures and tested only on restricted datasets. Our initial hypothesis is that combining a vision transformer with residual feature refinement can better exploit both global and local information in the system matrix, enabling the recovery of fine frequency components that are otherwise lost in low-resolution representations. We further hypothesize that this hybrid design will generalize across both experimental Open MPI data and simulated datasets with variable coil sensitivities, leading to more accurate system matrix recovery and improved image reconstruction compared to existing super-resolution approaches.

Unlike previous methods that either (i) rely on standalone architectures such as pure CNNs or Transformers, or (ii) neglect the variability of system matrix characteristics, our approach is to integrate vision transformers with residual feature refinement for MPI system matrix recovery and to validate this hybrid design on both experimental and simulated datasets. This dual evaluation highlights the model's ability to generalize beyond fixed scanner conditions and address practical challenges in MPI system calibration.

Innovative studies utilizing deep learning advancements have improved the MPI system matrix. For instance, Schrank *et al.* combined deep learning with local implicit image functions (LIIF), lowering recovery time by 90%, whereas dealing with high-frequency artifacts is challenging [18]. Baltruschat et al. introduced a 3D system matrix recovery network (3D-SMRnet), however, the method requires a system matrix calibrated for certain scan settings, particle types, and environmental conditions [19]. Gungor *et al.* created a transformer for system matrix super-resolution (TranSMS), but it required extensive training [17]. Shi *et al.* presented a progressive pretraining technique for high-resolution system matrix recovery (ProTSM), which enhanced performance by 15% but faced boundary artifacts [16]. Table 1 shows the key methodologies in MPI system matrix recovery from selected literature.

The following are our contributions to this work:
- Implementation of VRF-Net: we propose a hybrid deep learning model (VRF-Net) in which residual feature refinement is used to super-resolve the MPI system matrix on the Open MPI dataset using a paired-image super-resolution technique.
- Extension to variable system conditions: we provide a comprehensive explanation and evaluation of VRF-Net using a simulated MPI dataset that incorporates receiving coil sensitivity variations, in contrast to prior works that assume fixed system characteristics during training.
- Image reconstruction and benchmarking: we demonstrate image reconstruction using the recovered high-resolution system matrix and compare VRF-Net against state-of-the-art models to establish its effectiveness.

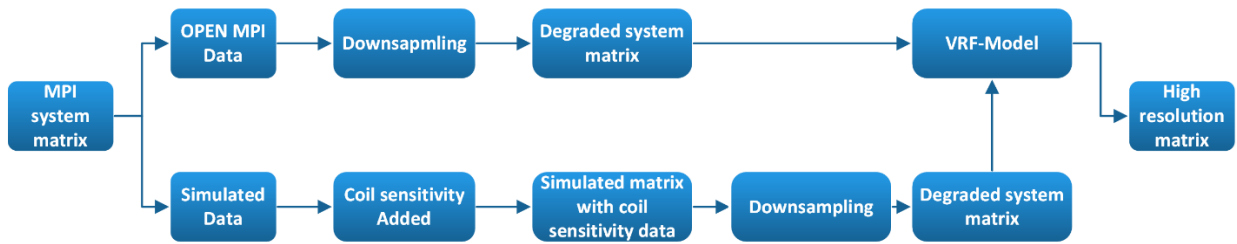

**Fig. 1.** Overview of the proposed method's workflow involves generating a degraded system matrix as a low-resolution input, followed by sequential downsampling. The ViT block and residual feature module

enhance resolution, reconstructing a high-resolution matrix with fine structural details and mitigating degradation artifacts.

**Table 1.** Comparison of key methodologies in MPI system matrix recovery and image reconstruction from selected literature. Checkmarks (✔) indicate features addressed in each study, while crosses (✘) denote

| Study | ViT using | Open MPI Dataset | Simulated Dataset | Coil Sensitivity Variation | Dual-Stage Downsampling | SM Recovery | Image Reconstruction | Multi-scale Eval. | Inference Time Analysis |
|---|---|---|---|---|---|---|---|---|---|
| **CCS**[21] | ✘ | ✘ | ✔ (2D Simulation) | ✘ | ✘ | ✔ (CS-based) | ✔ (SM-based) | ✘ | ✘ |
| **LIIF**[22] | ✘ (RDN) | ✔ | ✘ | ✘ | ✘ (Single-Stage) | ✔ | ✔ | ✘ (Fixed 16×) | ✘ |
| **DIP-SM**[23] | ✘ (CNN, U-Net-like) | ✔ | ✘ | ✘ | ✘ (Single-stage) | ✔ | ✔ | ✔ (2×,3×,4×) | ✔ (Slow) |
| **RETNet**[24] | ✔ (ResNet, DWConv) | ✘ | ✔ (MNIST, Phantoms) | ✘ | ✘ | ✘ | ✔ (X-space) | ✘ | ✘ |
| **MPI-GAN**[25] | ✘ cGAN + Decon) | ✔ (Limited) | ✔ | ✘ | ✘ | ✘ | ✔ (End-to-end) | ✘ | ✔ (Fast: 0.003s/image) |
| **DEQ-MPI**[26] | ✘ RDN) | ✘ | ✔ | ✘ | ✘ | ✘ | ✔ | ✘ | ✔ |
| **ProTSM**[16] | ✔ | ✔ | ✔ | ✘ | ✘ | ✔ | ✔ | ✔ | ✘ |
| **TranSMS**[17] | ✔ | ✔ | ✔ | ✘ | ✘ | ✔ | ✔ | ✔ | ✘ |
| **Multi-task TransGAN**[27] | ✔ (Transformer GAN) | ✔ | ✔ (Large-scale) | ✘ | ✘ | ✘ | ✔ (End-to-end) | ✘ | ✘ |
| **Ours** | ✔ (ViT-based) | ✔ | ✔ | ✔ (Simulated) | ✔ (Dual-stage) | ✔ | ✔ | ✔ (2×,4×,8×) | ✔ |

absent elements. The table highlights gaps in methods, datasets, and approaches across existing works.

## 2. Methods
### 2.1. Problem definition and approach

In this work, we consider the problem of recovering a high-resolution system matrix ($\hat{S}_{HR}$) from a degraded low-resolution counterpart $S_{LR} \in \mathbb{C}^{B \times C \times H \times W}$ with batch size $B$, channels $C$, width $W$, and height $H$, where the latter suffers from deteriorated frequency components ($\varepsilon_f$). The challenge lies in the high dimensionality of the target system matrix, $\hat{S}_{HR} \in \mathbb{R}^{N_X \times N_Y \times N_Z \times \varepsilon_f}$, where $N_X \times N_Y \times N_Z$ are the spatial dimensions require extensive sampling to accurately represent the mapping between the spatial distribution of MNPs and the detected MPI signal. To address this, we ask whether a hybrid deep learning framework, VRF-Net, which integrates global attention through vision transformers and local refinement through residual feature blocks, can effectively restore the degraded fine frequency components and reconstruct a system matrix with sufficient resolution to enhance downstream MPI image reconstruction. Furthermore, we investigate whether the recovered system matrices generalize well across both experimental Open MPI data and simulated datasets with varying coil sensitivity profiles.

### 2.2. VRF-Net Architecture
The model consists of a cascade of patch embedding blocks followed by a vision transformer, with the encoded output connected to the residual feature block. The input $S_{LR}$ is processed by the patch embedding block to extract spatial information and enable self-attention [28][29]. Increasing the number of patches enhances detail acquisition but may risk overfitting; hence, a convolutional layer with a kernel incorporates input channels $C_{in}$ and embedded dimension φ slides over $S_{LR}$, is utilized for adaptation[30].

To encode spatial information, a position embedding (PE) is added to the patch embedding, producing positionally encoded patch embeddings $\mathcal{X}_{PE}$, This output is processed by a multi-head self-attention (MHSA) block, in which query $Q_i$, key $K_i$, and value $V_i$ will be extracted through the softmax function from each patch as in Eq. (1). These projections enable the self-attention mechanism to compute how patches interact and aggregate information, helping the model reconstruct a high-resolution system matrix in both local and global contexts [31]. The MHSA block applied as:

$$MHSA_{(Q,K,V)} = softmax\left(\frac{(Q_i + \mathcal{X}_{PE}) \times K_i}{\sqrt{\varphi/\hbar}}\right) V_i \qquad (1)$$

where $\hbar$ is the number of heads in the MHSA block. The attention mechanism focuses on different regions of $S_{LR}$ based on the position embedding. The feature map then passes to a new feature space is a multi-layer perceptron (MLP), where it is first processed by a linear layer, followed by a ReLU activation, and then another linear layer to produce a new feature map $\mathcal{X}_{MLP}$ with a new number of channels $C_{out}$ as in Eq. (2). The MLP is coupled to two fully connected layers, fc1 and fc2, refining and modifying the features extracted by the transformer block. The denominator $\sqrt{\varphi/\hbar}$ serves as a normalization factor to prevent overly large dot-product values, stabilizing the softmax function, especially in deeper networks. The resulting attention matrix determines how much information from one patch should be passed to another, allowing the network to learn long-range dependencies, which are key in reconstructing fine structural details from degraded inputs.

$$\mathcal{X}_{MLP} = Linear_2(Linear_1((MHSA_{(Q,K,V)})) \qquad (2)$$

The MLP transforms the attention features into a higher-level representation, effectively allowing the model to perform feature extraction. In transformer architectures, this step enhances the model's expressiveness and enables non-linear combinations of the attended features. It also projects the feature map to a target dimension $C_{out}$, preparing it for further refinement. This transformation is essential for capturing composite and non-linear patterns in the system matrix, including subtle spatial correlations not easily modeled with attention alone.

To refine the features extracted by the transformer and recover high-frequency details that are often degraded during downsampling, the encoded output is passed into a residual feature block. Each residual block consists of two convolutional layers, each followed by batch normalization and a ReLU activation. These layers work together to emphasize subtle local structures, such as fine edges or frequency components that carry essential information for system matrix recovery. By incorporating skip connections, the block avoids vanishing gradients and allows the network to focus on learning the "corrections" needed to improve upon the coarse transformer features rather than relearning the entire mapping.

After feature refinement, the output is upsampled using a pixel-shuffling operation. This technique rearranges feature map elements from the channel dimension into the spatial dimension, effectively increasing resolution without introducing the blurring typically caused by interpolation. Concretely, a 2D convolution first increases the number of channels by a factor of $sf^2$, where $sf$ is the upsampling scale. Pixel shuffling then redistributes these channels into a higher-resolution grid, producing an output map $U_{(B,C_{out},H_k,W_k)}$ as in Eq.3. The kernel size $K$ and padding used in the convolution determine how much local context is blended during this expansion, ensuring smooth transitions and reducing artifacts.

$$U_{(B,C_{out},H_k,W_k)} = \mathcal{P}\left((Conv2d(X_{BN}, C_{in}\ sf^2, K, padding))\right) \tag{3}$$

Unlike conventional interpolation methods, pixel shuffling leverages learned feature representations rather than fixed formulas, making it especially powerful for tasks requiring fine structural accuracy. In our case, this enables the recovery of delicate system matrix patterns that directly impact the quality of reconstructed MPI images. Thus, while the transformer module captures global spatial relationships, the residual feature pathway ensures that local details are preserved and sharpened, providing the high-frequency fidelity necessary for accurate image reconstruction of the resolution phantom. While the transformer layers (MHSA and MLP) learn patch relationships and frequency patterns, the upsampling step ensures that all that learned information is properly mapped back to the spatial domain at a better resolution.

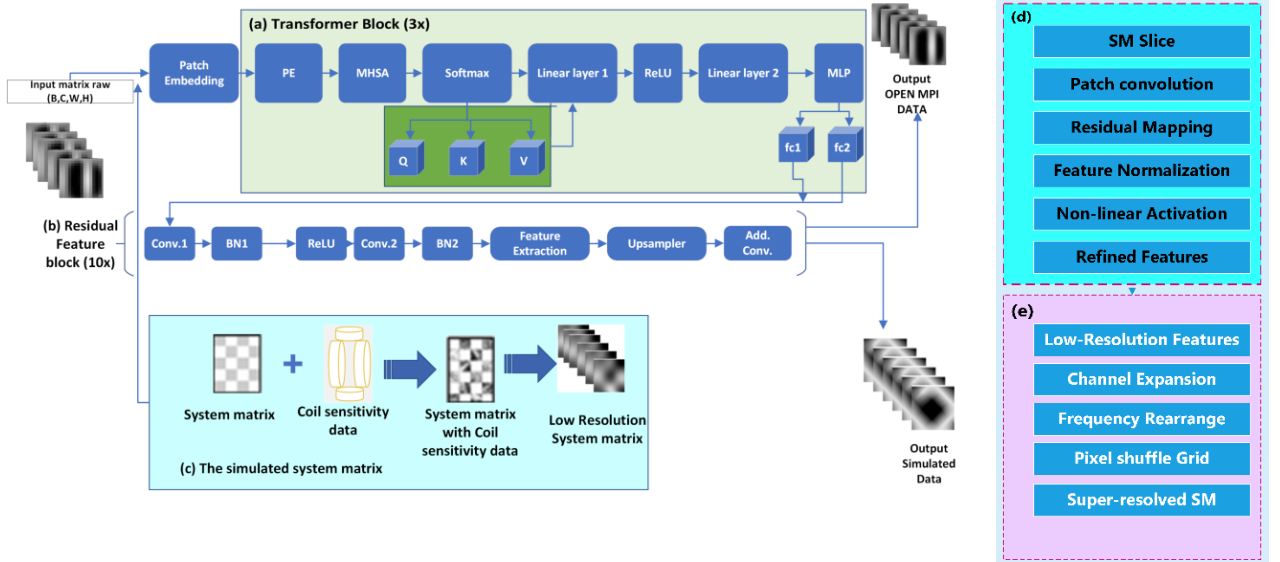

**Fig. 2.** (a) Schematic diagram of the proposed network architecture (VRF-Net) (stacked 3 times), the global features come from the transformer, and the localized features from the residual feature model (b) are combined and upsampled. The convolutional residual feature (stacked 10 times) module employs a feature extraction module with an upsampler to leverage the fine localized details in the input images by using the convolutional blocks. (c) The simulated system matrix with the coil sensitivity data. This generates a low-

resolution system matrix, where coil sensitivity effects are incorporated into the system matrix before degradation through downsampling. Figures (d) and (e) are the illustration feature extraction and upsampler blocks, respectively.

The overall output from the residual feature block $\mathcal{X}_f$ can be expressed in terms of $C_{in}, C_{out}, sf$, the number of features involved in $\mathcal{X}_{BN}$, which is denoted by (Ǹ), and $N_{resid.}$ denotes how many times the residual operation is applied, $\mathcal{X}_f$ is generated as:

$$\mathcal{X}_f = R(U_{(B,C_{out},H_k,W_k)}, C_{in}, C_{out}, sf, Ǹ, N_{resid.}) \tag{4}$$

here $C_{in}$ tells the residual block how many channels are coming into the block (i.e., the number of features it needs to analyze) and $C_{out}$ indicates how many channels the residual block will output after processing, shaping the feature map's depth after it passes through the residual block. Fig. 2 represents the architecture of the proposed network.

The residual block helps the network learn corrections to its predictions. Rather than trying to learn the entire mapping from degraded to high-resolution matrix in one go, the model learns to improve upon a coarse approximation iteratively.

In the feature extraction block, as shown in Fig.2 (d), a patch convolution applies local spectral filtering to highlight important patterns within each slice. A residual mapping path is then used as an error correction mechanism, ensuring that critical frequency components from the input are preserved. To stabilize learning, feature normalization is applied, scaling the frequency response across rows. The output then passes through a nonlinear activation, which amplifies small variations that carry subtle but important details. This stage also incorporates edge preservation, helping the model retain sharp transitions within the system matrix rows. Together, these steps yield Refined Features, where both fine spectral details and structural integrity are preserved.

In the upsampler block as shown in Fig. 2 (e), the low-resolution features produced from the residual path are first processed by channel expansion, which encodes hidden spectrum information by increasing the feature depth. This expanded representation undergoes frequency rearrange, mapping spectral channels into spatial positions. The pixel shuffle grid operation then upsamples the system matrix rows into a higher-resolution grid, reorganizing information without introducing blurring. Finally, the super-resolved version of the system matrix with enhanced fidelity is produced, which is then used for accurate image reconstruction of the resolution phantom.

### 2.3. VRF-Net configuration and training

The model's patch embedding module segments the input into 10×10 patches, embedding them into a 64-dimensional space with 64 output channels with a stride of 10. The transformer employs positional encoding to encode spatial information, helping to understand the spatial relationships between patches, which is crucial for image reconstruction.

The self-attention block employs multi-head attention with 8 heads, each with a dimension of 8, to capture long-range dependencies. The MLP block employs a hidden dimension of 128, with ReLU activation and linear transformations, ultimately shrinking the output to 64 channels.

To improve spatial resolution, the model incorporates 10 residual feature blocks, each containing 128 feature channels. These blocks employ 3×3 convolutional layers with a stride of 1 and utilize skip connections to maintain low-level characteristics. Upsampling is performed using scale factors of 2, 4, and 8, thereafter enhanced by additional convolutional layers for refinement [28]. The paired image super-resolution technique trains models on datasets containing corresponding low- and high-resolution image pairs $N$, allowing the model to learn features during training, in contrast to conventional methods that depend on fixed formulas and cannot adapt to the unique features of different datasets[32] [33] [34][35].

VRF-Net is trained using datasets with adaptive learning rate $\alpha_t$, and the loss function $L$ is computed as follows:

$$\min L(\hat{S}_{HR}, l_p, \varepsilon_f) = \frac{1}{N}\sum_{i=1}^{N}\|\hat{S}_{HR} - \text{VRFNet}(S_{LR}, l_p)\|^2 \tag{5}$$

$$l_{p_{t+1}} = l_p - \alpha_t \nabla L(\hat{S}_T, l_p, \varepsilon_f) \tag{6}$$

where $\hat{S}_T$ is the ground truth system matrix.

Training of the model involves a customized Mean Squared Error (MSE) loss function and optimization with the Adam optimizer using a learning rate between $10^{-3}$ and $10^{-7}$, $\alpha_t$ decayed over time by starting from $10^{-3}$ and decreasing towards $10^{-7}$ as the training progresses to stabilize convergence. Data augmentation methods, such as random horizontal flips, rotations, and cropping, were implemented. The values of mean and standard deviation, ranging from 0.05 to 1, respectively, were applied to the data for normalization. MSE is adopted to calculate the loss between $\hat{S}_{HR}$ predicted from the model and $\hat{S}_T$ incorporated with its feature map from each layer $\mathcal{X}_\varphi$ as:

$$L_{MSE}(\hat{S}_{HR}, \hat{S}_T) = \frac{1}{N}\sum_{i=1}^{N}(\mathcal{X}_\varphi(\hat{S}_{HR}(j)) - \mathcal{X}_\varphi(\hat{S}_T(j)))^2 \tag{7}$$

The model is trained for 350 epochs to recover the system matrix. All training was performed using the PyTorch framework on an Ubuntu 20.04.6 LTS operating system, with a system equipped with four GPUs.

### 3. Experimental design
### 3.1. Dataset
#### 3.1.1. Open MPI Dataset

An open-source dataset includes various imaging phantoms and calibration datasets containing the system matrices. The datasets taken are calibration data No.9 and No.10, which have the same grid size and FOV used as training data, whereas No.3 and No.8 are used as testing data. All the datasets used 3D Lissajous as a sequence with a drive field frequency of 2.5 MHz/102 × 2.5 MHz/96 × 2.5 MHz/99 and a selection field gradient of -0.1T/m×-1.0 T/m ×2.0 T/m. The resolution phantom consists of five tubes with 50 mmol of Perimag used to reconstruct the images, combined with the recovered system matrix. The data selected are 12,630 samples from the center of each slice, 9,744 for training, 1,886 for testing, and 1000 for validation. To integrate these 3D matrices into the model and manage the computational complexity of 3D system matrices, we adopted a dimensional decoupling strategy in which each 3D matrix was processed as a sequence of 2D slices. While this reduces memory and training requirements, care was taken to preserve inter-slice continuity. Specifically, each 2D slice encodes localized spectral patterns that remain consistent across neighboring slices because they originate from the same physical acquisition setup and frequency encoding. Furthermore, positional encoding within the vision transformer allows the model to capture contextual relationships within each slice, while residual feature learning ensures that fine-scale structures are consistently recovered. This slice-wise strategy is valid in MPI because each slice of the system matrix corresponds to independent spectral measurements defined by the field configuration. The inter-slice correlations are relatively weak compared to the strong intra-slice frequency encoding [36], meaning that most of the structural information is already contained within individual slices. Moreover, treating slices independently increases data diversity for training while significantly reducing computational load.

#### 3.1.2. Simulated dataset

The system matrix is initially simulated by modeling how magnetic nanoparticles respond to a time-varying magnetic field across a 3D spatial grid. At each grid point, the magnetization of the particles is computed based on their physical behavior using the Langevin function to capture how they align with the applied magnetic field over time. The imaging parameters in Table 2 are applied to generate the system matrix. Inspired by the fact that the correlation between the coil sensitivity and the spatial resolution of the system matrix is fundamental for enhancing the imaging performance of MPI [10], the data used here are simulated by applying different random values of coil sensitivities in x, y, and z directions. The parameters are adjusted to simulate the data represented in Table 2 for two data sets.

To demonstrate the coil sensitivities applied to the simulated system matrix, the time derivative of the magnetization under the influence of a magnetic field $\nabla \mathcal{M}(t)$ was analyzed, which can be mathematically given as:

$$\nabla \mathcal{M}(t) = \frac{-d}{dt} \int_V \mathcal{M}(\mathbf{r}, t) \left( \frac{\mu_0}{4\pi} \oint_{\partial S} \frac{\boldsymbol{\ell} - \mathbf{s}}{\|\boldsymbol{\ell} - \mathbf{s}\|_2^3} \times d\mathbf{l} d^3 \mathbf{r} \right) \tag{8}$$

where $\mathcal{M}(\mathbf{r}, t)$ is the magnetic moment density at position r and time t, $\mu_0$ is the permeability of free space, and $\ell$ and s are position vectors. The volume in Eq. (8) is conceptualized as a 3D grid, while the coil surface is characterized as a circular loop in the (x, y, z) plane with a designated radius; for every point within the 3D grid, the sensitivity arising from each specific point located on the surface of the coil is computed as the amount between the parentheses in Eq. (8). In the context of coil sensitivities, the sensitivity values in the x, y, and z dimensions $\eta_x$, $\eta_y$, and $\eta_z$ are calculated and result in sensitivity matrices $S_x$, $S_y$, and $S_z$. Knowing how the coils react to magnetic fields in various directions depends on these sensitivities, and for a given coil, the coil sensitivities in the x, y, and z directions are represented by each row in these matrices. Each sensitivity matrix has a dimension, and this results in $S_\sigma = (\eta_c, r_i)$ where $\sigma \in \{x, y, z\}$ representing the sensitivity of the coil ($\eta_c$) at grid point $r_i$ in the x, y, and z directions. Putting these sensitivity matrices together, a system matrix (A) with coil sensitivity data is generated as in Eq. (9). Fig. 5 represents the coil sensitivity profiles applied to produce the simulated data.

Each row in (A) is downsampled to reduce the resolution by selecting the maximum value within non-overlapping pooling windows along each row. The simulated data was generated using MATLAB (MathWorks, 23.2). Two datasets were produced, as illustrated in Table 2, each containing 9,000 samples, with 6,000 utilized for training, 2,000 for testing, and 1000 for validation.

The simulated system matrix (A) is given as:

$$A = \begin{bmatrix} S_x \\ S_y \\ S_z \end{bmatrix} = \begin{bmatrix} S_x(1,1) & S_x(1,2) & \cdots & S_x(1,N) \\ \vdots & \vdots & \ddots & \vdots \\ S_x(\eta_c,1) & S_x(\eta_c,2) & \cdots & S_x(\eta_c,N) \\ S_y(1,1) & S_x(1,2) & \cdots & S_x(1,N) \\ \vdots & \vdots & \ddots & \vdots \\ S_y(\eta_c,1) & S_x(\eta_c,2) & \cdots & S_x(\eta_c,N) \\ S_z(1,1) & S_z(1,2) & \cdots & S_y(1,N) \\ \vdots & \vdots & \ddots & \vdots \\ S_z(\eta_c,1) & S_y(\eta_c,2) & \cdots & S_y(\eta_c,N) \end{bmatrix} \tag{9}$$

The MATLAB pseudocode given in Algorithm 1 below provides a simulation of the coil sensitivity based on the Biot–Savart law, as expressed in Equation (8). The algorithm begins by initializing the simulation environment. Specifically, it sets up empty matrices $S_x$, $S_y$, and $S_z$ to store the sensitivity values in the x, y, and z directions, respectively. For each spatial voxel (i.e., point in the 3D field of view), the algorithm loops through all discrete segments that make up the circular receive coil. At each step, it calculates the vector representing the coil segment (dl) and the midpoint of that segment ($\ell$_mid). Then, it computes the vector between the segment midpoint and the voxel location (r_vec) and determines its magnitude. Using the Biot–Savart formulation, the magnetic field contribution of that coil segment at the voxel location is calculated and accumulated into a running total vector (B_total).

Once all coil segments have been considered for a given voxel, the x, y, and z components of the resulting magnetic field vector are stored in the corresponding sensitivity matrices: $S_x$, $S_y$, and $S_z$. This process is repeated for every voxel in the grid, thereby constructing a complete sensitivity profile of the coil across space.

**Algorithm 1: Coil Sensitivity Simulation**

function simulate_sensitivity (grid, coil_loop);
**Input:**
   - grid: 3D spatial points r_i arranged as (X × Y × Z)
   - coil_loop: Discretized positions ℓ on the coil surface
**Output:**
   - $S_x$, $S_y$, $S_z$: Sensitivity matrices in x, y, z directions

1: Initialize $S_x$, $S_y$, and $S_z$ to zero matrices of size (num_coils × num_voxels)
2: μ₀ ← 4π × 10⁻⁷    → Permeability of free space
3: for each spatial point r in grid do
4:    B_total ← (0, 0, 0)    → Initialize magnetic field at r
5:    for each segment j on coil_loop do
6:      dl ← ℓ_{j+1} − ℓ_j    → Vector along coil segment
7:      ℓ_mid ← (ℓ_{j+1} + ℓ_j) / 2    → Midpoint of the coil segment
8:      r_vec ← r − ℓ_mid    → Vector from coil to voxel
9:      r_mag ← norm(r_vec)
10:     if r_mag ≠ 0 then
11:       dB ← (μ₀ / (4π)) × (dl × r_vec) / r_mag³    → Biot–Savart
12:       B_total ← B_total + dB
13:     end if
14:    end for
15:    $S_x$(: r_index) ← B_total_x
16:    $S_y$(: r_index) ← B_total_y
17:    $S_z$(: r_index) ← B_total_z
18: end for
19: return ($S_x$, $S_y$, $S_z$)
end function

### 3.2. Dataset preprocessing

To manage the complex frequency components of the system matrix for both types of data, we preprocess the data by dividing the complex values into their real and imaginary parts. We treat each row of the system matrix independently. Each row corresponds to a distinct spatial frequency, which offers localized spectrum information. By processing rows independently, we focus on capturing the fine-grained features specific to each frequency component while eliminating needless interactions between unrelated frequencies. Furthermore, the real and imaginary components of each row are normalized independently to standardize their sizes. The system matrix data for the Open MPI dataset is generated with MATLAB (MathWorks, 23.2) and selected with an SNR threshold ≥ 10, while keeping the simulated data with its original noise level[37]. Next, the data undergoes a resolution reduction employing max-pooling to reduce the resolution of each system matrix row. Once again, the rows are downsampled by average pooling before being fed to the model, and the resulting low-resolution system rows are then reshaped to meet the model's dimensions. The training and testing data for the simulated dataset were collected across a range of coil sensitivity values from 0.5 to 1 mV/μT. This results in avoiding spatial encoding errors and achieving a consistent sensitivity profile across a specified FOV.

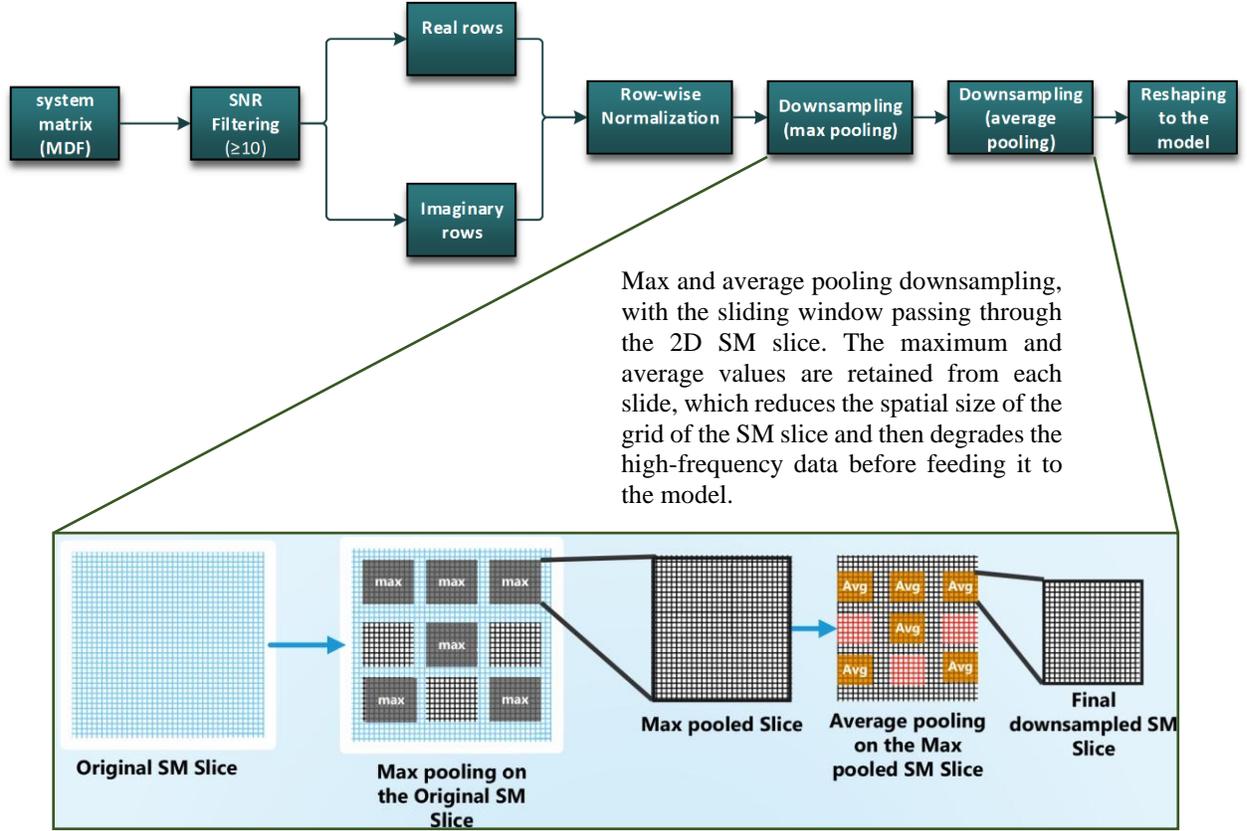

**Fig. 3.** The block diagram of the preprocessing pipeline: the system matrix data is preprocessed first into the magnetic particle imaging data format (MDF) for the Open data before being fed to the model. The dual downsampling process is also given in the magnified diagram below the main diagram, the highlighted small grid in gray for the max pooling and orange in the average pooling are approximations that represent the selected values according to the window stride value. The preprocessing for the simulated data is the same as the Open data, except for the SNR filtering.

**Table 2.** The parameters used to simulate the system matrix.

| Parameter | Value |
|---|---|
| **Number of coils** | 3 |
| **FOV (mm)** | [30×30×30] and [45×45×45] |
| **Sampling frequency (Hz)** | $10^6$ for both |
| **Drive frequency (Hz)** | [26e3, 25e3, 27e3] and [30e3, 36e3, 39e3] |
| **Particle size (nm)** | 15.7 and 25 |
| **Selection field gradient (T/m)** | -0.1 ×-1.0 ×2.0 for both |
| **Drive field amplitude (mT)** | 12 × 12 × 12 |

### 3.3. Benchmarks and evaluation metrics
#### 3.3.1. Bicubic interpolation
Is the most common super-resolution procedure used in 2D images, here it comprises three convolutional layers, and each layer applies several filters of size 9×9 and 1×1 with 4- and 2 pixel padding to preserve spatial dimensions [38].

#### 3.3.2. SRCNN
Super-Resolution Convolutional Neural Network is the basic method for recovering a high-resolution image from a single low-resolution image empowered by CNN with lightweight architecture. For this work, SRCNN comprises three convolutional layers; the first is responsible for capturing extensive context regions, the second is dedicated to transforming feature maps, and the last is focused on generating a single-channel image [39].

#### 3.3.3. VDSR
Very Deep Super-Resolution network; it comes to fix the limitations of SRCNN, here it consists of 18 residual layers, where each block is comprised of a convolutional layer followed by a ReLU activation function [40].

#### 3.3.4. MDSR
Multi-scale super-resolution network is an enhanced deep super-resolution network (EDSR) advancement. It consists of a cascade of residual blocks and convolutional layers, followed by an upsampling model [41].

#### 3.3.5. TranSMS
Transformer for Super-resolution System Matrix, an innovative transformer application in optimizing MPI calibration. The model consists of a transformer, a convolutional, and a data consistency module[17].

#### 3.3.6. ProTSM
A novel approach for fast 3D system matrix calibration using the Progressive Pretraining Network mechanism. This method allows the model to efficiently use unlabeled low-resolution SM data to prevent overfitting and boost performance when labeled data is limited. This method is only adopted in image reconstruction results [16].

To assess the performance of the VRF-Net compared to these approaches in system matrix recovery and image reconstruction, the peak signal-to-noise ratio (pSNR), structural similarity index (SSIM), and normalized mean square error (nRMSE) were calculated for each method at different scale factors, which are standard and widely adopted in image super-resolution and MPI reconstruction studies. These metrics allow direct comparison with prior MPI-related work and provide complementary views of fidelity. For all competing methods, the input was the same as that used for the VRF-Net. All these metrics are computed as follows:

$$pSNR = 20.\log_{10}\left(\frac{\left(max(S_{ref})\right)^2}{\frac{1}{N}\sum_{i=1}^{N}\left(S_{rec}(i)-S_{ref}(i)\right)^2}\right) \quad (10)$$

where $S_{ref}$, $S_{rec}$ and N denote the reference, recovered system matrix slice, and the total number of pixels in the slice, respectively. The nRMSE was calculated as:

$$nRMSE = \frac{\sqrt{\frac{1}{N}\sum_{i=1}^{N}\left(S_{rec}(i)-S_{ref}(i)\right)^2}}{max(S_{ref})-min(S_{ref})} \quad (11)$$

which normalizes the recovery error relative to the dynamic range of the reference system matrix slice. SSIM was used to capture structural fidelity and is defined as:

$$SSIM = \frac{(2\mu_x\mu_y + C_1)(2\sigma_{xy} + C_2)}{(\mu_x^2+\mu_y^2+C_1)(\sigma_x^2+\sigma_y^2+C_2)} \quad (12)$$

Where $\mu, \sigma^2$ and $\delta\sigma_{xy}$ represent local means, variances, and covariance, respectively, while $C_1$ and $C_2$ are small constants to stabilize the division when denominators are close to zero.

**Table 3.** Parameters used in this work.

| Parameter | Unit | Symbol | Description |
|---|---|---|---|
| SM | | System Matrix | The system matrix term is used partially in this work. |
| $S_{LR}$ | - | Low-resolution system matrix | Input SM with degraded frequency components |
| $\hat{S}_{HR}$ | - | High-resolution system matrix | Super-resolved system matrix recovered by VRF-Net |
| B | - | Batch size | Number of training samples in each iteration |
| C | - | Channels | Number of feature channels in input tensor |
| H, W | pixels | Height and Width | Spatial dimensions of the SM slice |
| $\varepsilon_f$ | - | Frequency components | Lost/deteriorated frequency components |
| $N_X, N_Y, N_Z$ | voxels | Spatial dimensions | Size of reconstructed MPI volume along X, Y, Z |
| $C_{in}$ | - | Input channels | Number of channels entering a Conv layer |
| $C_{out}$ | - | Output channels | Number of channels after Conv/MLP |
| $\varphi$ | - | Embedding dimension | Latent dimension used in patch embedding |
| PE | - | Positional embedding | Encodes spatial position of patches for transformer |
| $Q_i, K_i, V_i$ | - | Query, Key, Value | Projection vectors in the multi-head self-attention (MHSA) block |
| $\hbar$ | - | Number of heads | Number of attention heads in MHSA |
| $sf$ | - | Upsampling scale factor | Factor used in pixel shuffle upsampling (e.g., 2, 4, 8) |
| $K$ | - | Kernel size | Size of convolution kernel in residual/upsampling blocks |
| $U_{(B,C_{out},H_k,W_k)}$ | - | Upsampled output map | High-resolution feature map after pixel shuffle |
| $H_k, W_k$ | pixels | Upsampled height and width | Spatial dimensions after pixel shuffle |

| Symbol | Units | Name | Description |
|---|---|---|---|
| $x_{PE}$ | - | Positionally encoded patch embedding | Patch embedding + positional encoding |
| $x_{MLP}$ | - | Feature map after MLP | Refined output of transformer block |
| $x_{BN}$ | - | Batch normalized features | Intermediate features after BN |
| $x_f$ | - | Residual block output | Output feature map from residual block |
| $N_{resid.}$ | - | Number of residual blocks | How many times the residual operation is applied |
| $\grave{N}$ | - | Number of BN features | Features involved in batch normalization within residual block |
| $x_\varphi$ | - | Feature map from the φ layer | Feature representation associated with embedding dimension φ |
| $\alpha_t$ | - | Adaptive learning rate | Learning rate at training step t (decays from 1e-3 to 1e-7) |
| $L$ | - | Loss function | Objective function minimized during training (MSE variant) |
| $\hat{S}_T$ | - | Ground truth SM | Reference high-resolution system matrix used for supervision |
| N | - | Training pairs | Number of training image pairs used in paired super-resolution |
| $\mu_o$ | H/m | Permeability of free space | Physical constant in coil sensitivity simulation ($4\pi \times 10^{-7}$ H/m) |
| $\eta_x, \eta_y, \eta_z$ | mV/μT | Coil sensitivities | Sensitivity values in x, y, z directions |
| $S_x, S_y, S_z$ | - | Sensitivity matrices | Coil sensitivity matrices for x, y, z directions |
| A | - | Simulated system matrix | System matrix incorporating coil sensitivities |
| r | m | Position vector | Spatial point location in coil sensitivity simulation |
| ℓ, ℓ_mid | m | Coil segment vectors | Discretized coil position and midpoint vectors |
| r_vec | m | Voxel-coil vector | Vector from coil segment midpoint to voxel position |
| r_mag | m | Vector magnitude | Norm of r_vec used in Biot–Savart calculation |
| B_total | T | Magnetic field vector | Accumulated magnetic field at the voxel due to the coil segments |

## 4. Results
### 4.1. System matrix recovery
#### 4.1.1. For the Open MPI dataset

The VRF-Net is demonstrated on Open MPI data, as in Table 4. The bicubic interpolation demonstrates the weakest performance in recovering the system matrix, exhibiting the largest nRMSE and the lowest SSIM across all scaling factors. Although SRCNN offers certain enhancements, its efficacy is still lacking, exhibiting significant errors and inadequate detail restoration, especially at higher scaling factors. VDSR has inferior performance relative to SRCNN, generating excessive artifacts and providing minimal enhancement in image quality. MDSR improves upon existing models by utilizing multi-scale feature extraction, leading to clearer images, fewer errors, and enhanced SSIM. TranSMS and VRF-Net yield remarkable outcomes, with TranSMS substantially reducing errors and more adeptly recovering intricate details compared to previous models. VRF-Net surpasses all alternative methods, attaining minimal recovery errors and yielding the most precise and visually coherent outcomes. Despite higher scaling factors, VRF-Net sustains its performance by retrieving fine features and maintaining structural integrity with few artifacts.

From Fig. 4, in terms of image quality, the Bicubic interpolation creates blurry images with poor detail preservation across all cases; it was unable to capture fine structures, resulting in large inaccuracies, particularly in high-frequency regions. SRCNN improves upon Bicubic with more precise images but introduces artifacts and problems with restoring details, as seen by its variable intensity and noticeable faults in complicated structures. VDSR performs better than SRCNN, yielding significantly less distorted representations with less fragmented information and fewer errors. MDSR exhibits great improvement, restoring clearer and more consistent details. Whereas it manages to regain structural correctness better than Bicubic, SRCNN, and VDSR, slight blurring and subtle artifacts persist, particularly in high-frequency regions.

TranSMS considerably outperforms previous CNN-based algorithms, giving more precise images with well-preserved high-frequency features and few artifacts. Its reconstructions maintain accuracy across different frequencies, with dramatically decreased error regions. VRF-Net outperforms TranSMS, delivering recovery that closely approximates the ground truth. It maintains subtle details and structural fidelity even at the highest frequencies, as demonstrated in its clean error maps and sharp restorations.

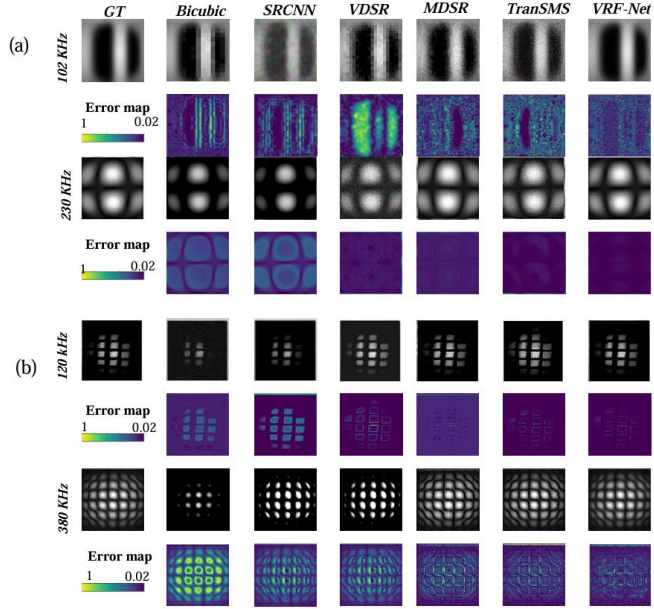

**Fig. 4.** The recovered system matrix compared with the ground truth (GT) with its corresponding error map for the Open MPI dataset at different frequencies, (a) the images taken for a scale factor of 2× at the center of the slice, whereas in (b) the images were taken for a scale factor of 4× at the center of the slice.

**Table 4.** Quantitative metrics for the system matrix recovery on the Open MPI dataset, FOV, and grid size are different for training data.

| Factor | 2× | | | 4× | | | 8× | | |
| --- | --- | --- | --- | --- | --- | --- | --- | --- | --- |
| Metrics | nRMSE | pSNR (dB) | SSIM | nRMSE | pSNR (dB) | SSIM | nRMSE | pSNR (dB) | SSIM |
| Method | | | | | | | | | |
| Bicubic | 47.186 | 19.50 | 0.420 | 78.174 | 15.72 | 0.417 | 137.18 | 13.39 | 0.409 |
| SRCNN | 41.06 | 19.92 | 0.508 | 34.89 | 17.95 | 0.620 | 136.05 | 15.66 | 0.609 |
| VDSR | 10.05 | 21.25 | 0.589 | 26.05 | 19.53 | 0.641 | 121.27 | 16.27 | 0.634 |
| MDSR | 9.94 | 29.10 | 0.647 | 17.78 | 26.90 | 0.678 | 58.19 | 21.86 | 0.598 |
| TranSMS[a] | 3.15 | 35.39 | 0.797 | 6.19 | 34.15 | 0.684 | **20.58** | 30.80 | 0.704 |
| VRF-Net | **0.403** | **39.08** | **0.835** | **3.404** | **38.06** | **0.729** | 46.404 | **31.06** | **0.717** |

[a]pSNR and SSIM for TranSMS were calculated based on their code, available on GitHub at https://github.com/icon-lab/TranSMS .

### 4.1.2. For the simulated data

The dataset is utilized to explore various FOVs, applied coil sensitivity values, and MNP sizes, as detailed in Table 2, however, we focused on the coil sensitivity variations. From Table 5. The VRF-Net attains an average enhancement of 88.2% in nRMSE, 44.7% in pSNR, and 34.3% in SSIM relative to Bicubic, SRCNN, VDSR, MDSR, and TranSMS. Fig. 6 illustrates the influence of coil sensitivity variations on system matrix recovery. Bicubic interpolation yields excessively blurred outcomes, inadequately preserving intricate details and generating considerable inaccuracies across all frequencies. SRCNN and VDSR, exhibit certain enhancements, with VDSR demonstrating superior capability in maintaining structural features. However, both struggle with recovering high-frequency components accurately, leading to artifacts and errors. MDSR exhibits superior performance, particularly at lower frequencies; nevertheless, it encounters difficulties with deterioration at higher frequencies due to the complexity of the system matrix. Transformer-based models, TranSMS, outperform CNN-based approaches in capturing high-frequency information and minimizing artifacts. Among all methodologies, VRF-Net delivers recovery nearly matching the ground truth across all frequency ranges. These changes in the structure of the system matrix, as in Fig. 6, result from the different sensitivity of the receiving coils applied in x and y directions, whereas it was constant along the z-direction, as in Fig.5, this alters the spatial distribution and amplitude of the received signals. The system matrix produces unique patterns at different frequencies, driven by coil sensitivity. The patterns are smoother and less complicated at lower frequencies (71 kHz), whereas higher frequencies (290 kHz) bring intricate structures due to the combined effects of increasing spatial resolution and coil sensitivity. These sensitivity-induced fluctuations pose challenges for accurate system matrix recovery since they produce non-linear distortions that models must adapt to for precise recovery.

**Table 5.** Quantitative metrics for the system matrix recovery on the simulated data. FOV, MNP size, and coil sensitivity values were simulated for training and testing data

| Factor | 23 | | | 4× | | | 8× | | |
|---|---|---|---|---|---|---|---|---|---|
| Metrics<br>Method | *nRMSE* | *pSNR (dB)* | *SSIM* | *nRMSE* | *pSNR (dB)* | *SSIM* | *nRMSE* | *pSNR (dB)* | *SSIM* |
| Bicubic | 45.05 | 18.38 | 0.543 | 69.02 | 15.11 | 0.435 | 89.05 | 13.88 | 0.408 |
| SRCNN | 34.31 | 20.20 | 0.567 | 55.38 | 17.42 | 0.539 | 79.45 | 15.28 | 0.491 |
| VDSR | 31.12 | 24.83 | 0.543 | 41.20 | 20.51 | 0.493 | 71.16 | 17.18 | 0.402 |
| MDSR | 18.25 | 25.76 | 0.660 | 14.20 | 21.30 | 0.573 | 34.23 | 19.30 | 0.489 |
| TranSMS | **4.33** | 27.24 | 0.691 | 8.36 | 23.87 | 0.671 | 11.06 | 22.76 | 0.567 |
| VRF-Net | 4.44 | **28.52** | **0.771** | **6.28** | **26.91** | **0.701** | **11.01** | **23.34** | **0.600** |

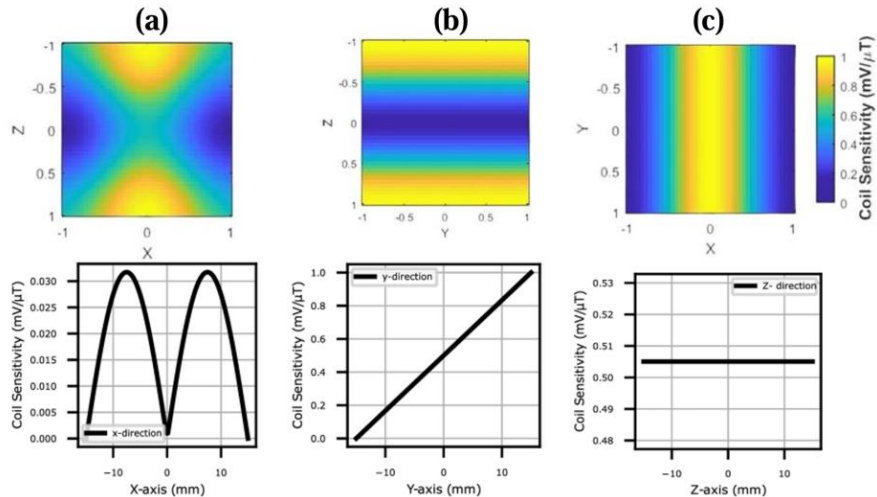

**Fig. 5.** The applied coil sensitivity (in mV/μT) matrix appeared as a spatially dependent profile as in [11] over the system matrix on the (a) x-direction, (b) y-direction, and (c) z-direction. Each profile shows, along with its coil sensitivity variation plot, in each direction.

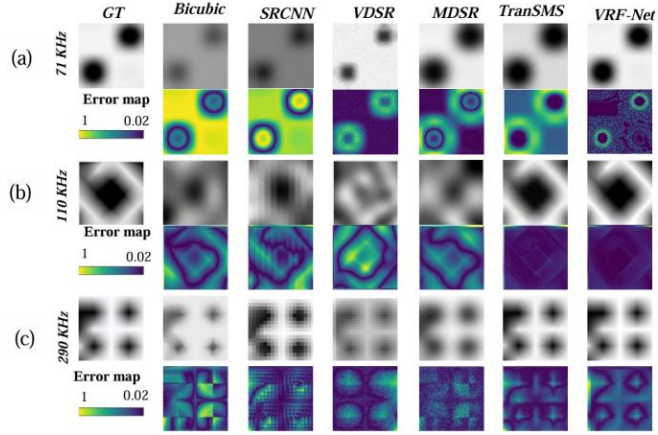

**Fig. 6.** The recovered system matrix compared with the ground truth (GT) with its corresponding error map for the simulated dataset, the images taken for a scale factor of 2× at the center of the slice, for an FOV of 30mm ×30mm ×30mm and different frequencies (a) 71 kHz, (b)110 kHz, and (c)290 kHz.

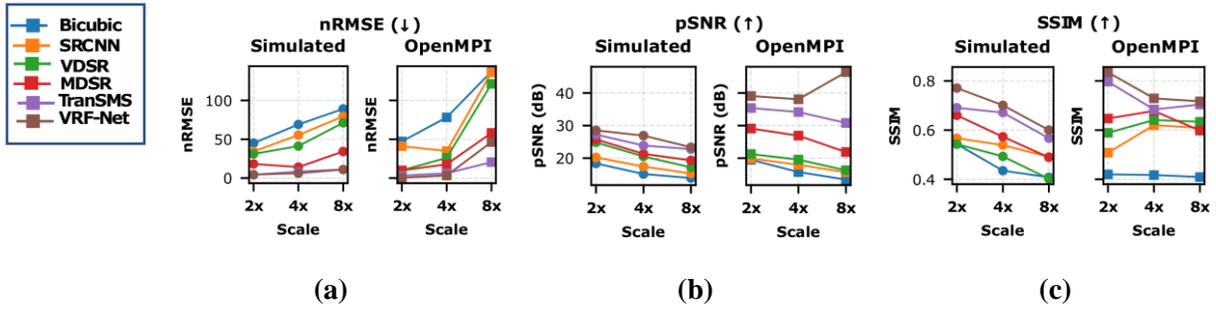

**(a)**        **(b)**        **(c)**

**Fig.7.** summarized metrics results for the system matrix recovery of the simulated and the Open MPI Data, (a) for the nRMSE (lower better), (b) pSNR (higher better), and (c) SSIM (higher better).

To visually assess how VRF-Net learns to separate low- and high-resolution features, we applied t-distributed Stochastic Neighbor Embedding (t-SNE) to the patch-wise feature embeddings. Four plots are presented as shown in Fig.8: training and testing feature distributions for both the Open MPI dataset and the simulated dataset. Each plot shows the low-resolution input features (LR) and their corresponding high-resolution reconstructions (HR) in two-dimensional space. The t-SNE visualizations demonstrate that VRF-Net effectively transforms LR features toward distinct HR representations while preserving intra-class structure. These plots provide qualitative confirmation that the network consistently captures meaningful patterns in both training and testing data, complementing the obtained quantitative results.

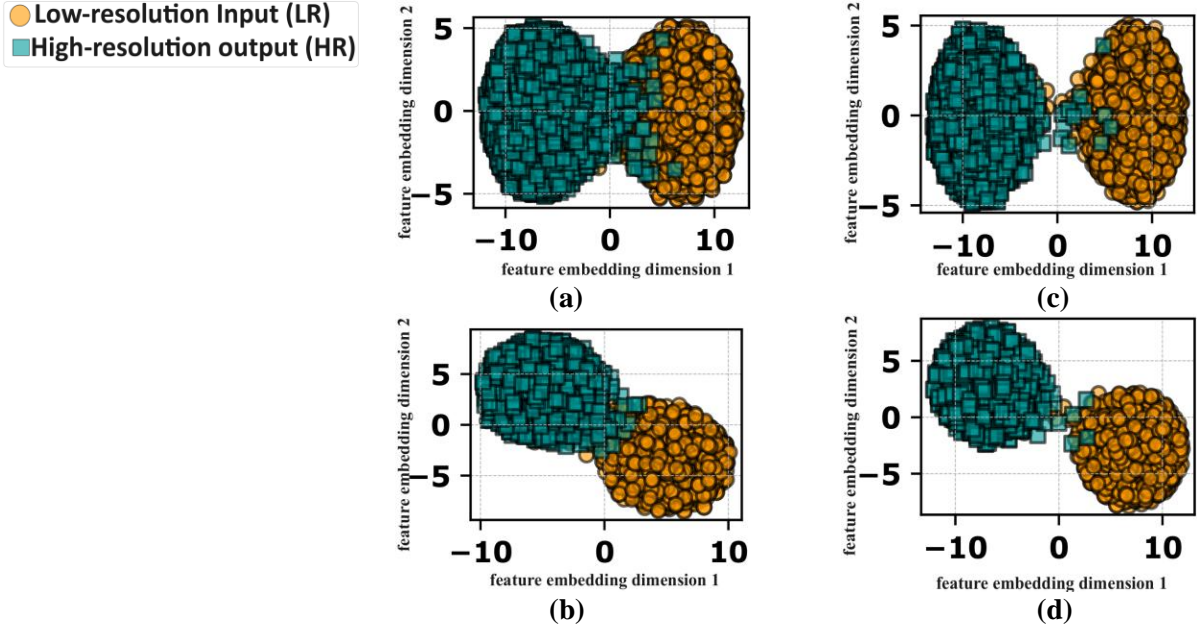

**Fig. 8.** The t-SNE plots for the training and testing features for both datasets used in this study: (a) Training features of the Open Data, (b) Testing features of the Open Data (c) Training features of the simulated data, and (d) Testing features of the simulated data. The plots visualize the high-dimensional feature embeddings learned by the VRF-Net, projected into a 2D space using t-SNE. Orange points represent features from low-resolution (LR) input features, while green points show features from the corresponding high-resolution (HR) output. The plots demonstrate the model's ability to learn and cluster distinct features during training and testing using both datasets. Overlapping regions suggest that, in the learned feature space, certain LR and HR features are sufficiently similar for successful recoveries.

### 4.2. Image reconstruction

To evaluate the quality of the recovered system matrix, we performed image reconstruction using a resolution phantom from the Open MPI dataset. This phantom provides a known structural target that helps assess how well fine details are preserved during the reconstruction process. The images were reconstructed using the Kaczmarz method, an iterative algorithm commonly used in MPI for solving large systems of linear equations. In our case, it was applied to estimate the image by solving for the unknown signal distribution using the recovered high-resolution system matrix $\hat{S}_{HR}$ and the corresponding measurement vector from the resolution phantom. To stabilize the reconstruction and avoid overfitting, we introduced regularization, which penalizes extreme solutions and ensures a more robust result. The regularization strength, denoted by $\lambda$, was set to the product of 1e-3 and the Frobenius norm of the system matrix, yielding a value of $\lambda= 0.69$ during the first iteration. The reconstruction process involved three sequential updates, progressively refining the solution to produce a sharper, more accurate image.

As shown in Table 6, the proposed VRF-Net outperforms the closest competing models, including TranSMS and ProTSM, in terms of reconstruction accuracy. Specifically, VRF-Net reduces the nRMSE by an average of 41.4%, meaning it produces significantly fewer errors when reconstructing the system matrix. In terms of pSNR, VRF-Net also leads with a 1.5% average improvement over ProTSM, indicating better preservation of high-intensity details. Additionally, the SSIM improves by 1.1% on average, confirming that VRF-Net preserves structural features more effectively.

It's also worth noting that performance metrics for 8× scaling were not reported for ProTSM in[16], which limits a complete comparison at higher resolution levels. Among the other methods, Bicubic interpolation consistently performs the worst across all scales, showing the highest reconstruction nRMSE and the lowest image quality metrics (pSNR and SSIM). SRCNN and VDSR provide modest improvements over Bicubic, with VDSR reliably outperforming SRCNN at each scale.

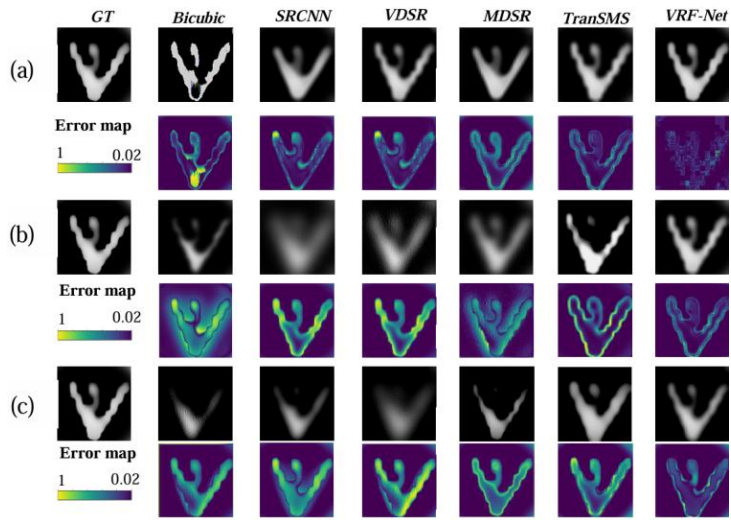

**Fig. 9.** The reconstructed images are compared with the GT image, competing methods, and error maps. (a) The reconstructed images at 2×, (b) at 4×, and (c) at 8×.

**Table 6.** Quantitative evaluation of the reconstructed images using the Open MPI data compared with other State-of-the-art methods

| Factor | 2× | | | 4× | | | 8× | | |
|---|---|---|---|---|---|---|---|---|---|
| Metrics | *nRMSE* | *pSNR (dB)* | *SSIM* | *nRMSE* | *pSNR (dB)* | *SSIM* | *nRMSE* | *pSNR (dB)* | *SSIM* |
| **Method** | | | | | | | | | |
| Bicubic | 38.06 | 20.76 | 0.410 | 67.04 | 16.99 | 0.407 | 98.04 | 14.02 | 0.400 |
| SRCNN | 35.23 | 21.24 | 0.548 | 32.24 | 20.24 | 0.511 | 91.29 | 17.60 | 0.477 |
| VDSR | 13.10 | 26.00 | 0.648 | 23.08 | 19.79 | 0.597 | 79.08 | 18.54 | 0.535 |
| MDSR | 7.46 | 28.37 | 0.656 | 15.17 | 24.39 | 0.591 | 43.30 | 21.90 | 0.540 |
| TranSMS[b] | 3.32 | 38.54 | 0.738 | 10.66 | 31.96 | 0.610 | 114.45 | 13.38 | 0.603 |
| ProTSM | **0.86** | 41.43 | 0.941 | 2.13 | 33.34 | 0.737 | - | - | - |
| VRF-Net | 1.79 | **41.58** | **0.960** | **2.09** | **34.74** | **0.746** | **32.80** | **32.26** | **0.633** |

[b]SSIM values were calculated based on their code, which is available on GitHub at https://github.com/icon-lab/TranSMS .

**Table 7.** A comparison of reconstructed image quality in Fig.9 and visual fidelity across all methods with their corresponding representative metric values at 2× and 4× scale factors.

| Method | Visual fidelity (qualitative) | Key artifacts / Errors | Representative values |
|---|---|---|---|
| **Bicubic** | -Blurry, loss of fine details.<br>-Poor edge preservation. | Widespread errors in high-frequency regions (bright patches in error maps). | -nRMSE: 38.06 (2×), 67.04 (4×).<br>-SSIM: 0.548 (2×), 0.407 (4×). |
| **SRCNN** | - Moderate improvement over Bicubic but still soft.<br>-Struggles with high-frequency features. | Residual blurring and incomplete detail recovery. | -nRMSE: 35.23 (2×), 32.24 (4×).<br>-SSIM: 0.548 (2×), 0.511 (4×). |
| **VDSR** | - Sharper than SRCNN but with minor artifacts.<br>-Better structural retention but limited global context. | Edge distortions and faint ghosting artifacts | - nRMSE: 13.10 (2×), 23.08 (4×).<br>- SSIM: 0.648 (2×), 0.597 (4×). |
| **MDSR** | - Improved clarity and multi-scale feature handling.<br>- Balanced but not optimal for high-frequency recovery. | -Slight blurring in complex regions.<br>-Fewer errors but uneven distribution. | - nRMSE: 7.46 (2×), 15.17 (4×).<br>- SSIM: 0.647 (2×), 0.678 (4×). |
| **TranSMS** | -High-frequency details are better preserved than CNNs.<br>- Transformer benefits are visible but lack local refinement. | Sparse bright errors (bright spots in error maps) | - nRMSE: 3.32 (2×), 10.66 (4×).<br>- SSIM: 0.738 (2×), 0.610 (4×). |
| **VRF-Net** | -Closest to GT with sharp edges and minimal artifacts.<br>-Best preservation of fine structures and textures. | -Negligible errors (darkest error maps).<br>-No boundary distortions or blurring. | - nRMSE: 1.79 (2×), 2.09 (4×).<br>-SSIM: 0.960 (2×), 0.746 (4×). |

Among all the baseline models, TranSMS is the closest in performance to VRF-Net in terms of quantitative metrics. However, visual analysis reveals clear qualitative differences.

At the 4× scale, both models succeed in recovering the overall structure. Still, VRF-Net demonstrates noticeably sharper edges and more precise localization of branches, as highlighted by the red boxes in Fig.10. These regions emphasize VRF-Net's ability to preserve fine structural features, particularly at the tips and intersections better than TranSMS.

At the more challenging 8× scale, the gap becomes more evident. TranSMS struggles with maintaining structural integrity, leading to visible blurring and loss of detail in the upper branches. In contrast, VRF-Net manages to retain both the shape and symmetry of the original signal, showing strong robustness even under severe downsampling.

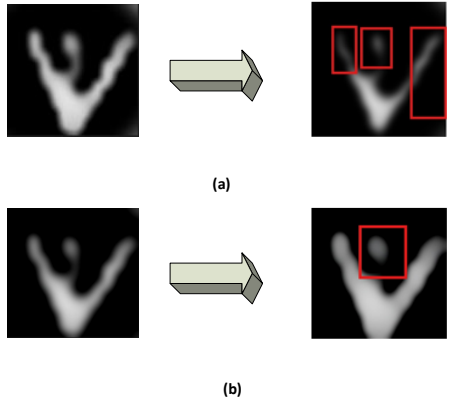

**Fig.10.** The difference in reconstructed structures results from the VRF-Net (left column) and the TranSMS (right column) for scale factors of 4× (a) and 8× (b).

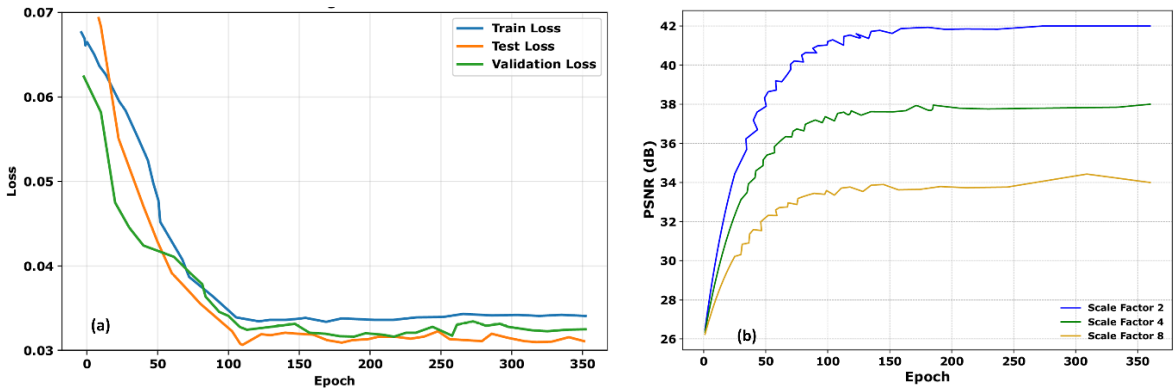

**Fig.11.** The VRF-Net performance in image reconstruction over the number of epochs on the Open MPI dataset, (a) the learning curve at a scale factor of 2×, and (b) the pSNR for different scale factors used.

### 4.3. Ablation study
### 4.3.1. On different model architectures

We investigated how different model architectures influence the recovery of the system matrix, comparing VRF-Net with several models, namely: a Vision Transformer with Residual Blocks (ViT-RB), a Convolutional Network-based Vision Transformer (ViT-CNN), and a Vision Transformer integrated with an upsampling mechanism (ViT-only). Table 8 shows that VRF-Net surpassed the other models and improved the recovered system matrix. We apply the ablation to the simulated dataset.

From Table 8, at the 2× scale, where the degradation is moderate, VRF-Net achieves the lowest normalized root mean square error (nRMSE = 4.44), the highest peak signal-to-noise ratio (pSNR = 28.52 dB), and the highest structural similarity index (SSIM = 0.771). In contrast, ViT-RB, while incorporating residual features, lags with a higher error (nRMSE = 6.97) and a significantly lower SSIM of 0.619, highlighting the importance of the full integration of both global and local feature extraction.

ViT-CNN and ViT-only perform worse, particularly in preserving structural fidelity. ViT-CNN shows an nRMSE of 14.01 and SSIM of just 0.453, indicating that convolution alone without residual learning or transformer refinement is insufficient. Similarly, ViT-only lacks the local detail modeling needed for accurate recovery, with an SSIM of 0.541.

At the more challenging 4× scale, the differences become even more noticeable. VRF-Net still performs strongly (nRMSE = 6.28, pSNR = 26.91 dB, SSIM = 0.701), while the other models show significant degradation. ViT-only in particular experiences a sharp drop in pSNR (20.53 dB) and a high nRMSE of 17.23, indicating that global modeling alone cannot make up for aggressive downsampling. ViT-CNN performs slightly better but still struggles to restore detailed structures (SSIM = 0.409).

These results confirm that each component of VRF-Net plays a vital role, the residual feature path ensures local detail preservation, while the transformer-based global modeling enhances structural coherence.

**Table 8.** The ablation study on the simulated dataset at scale factors of 2× and 4×

| Factor | 2× | | | 4× | | |
|---|---|---|---|---|---|---|
| Metrics | *nRMSE* | *pSNR (dB)* | *SSIM* | *nRMSE* | *pSNR (dB)* | *SSIM* |
| **Method** | | | | | | |
| ViT-RB | 6.97 | 27.78 | 0.619 | 8.76 | 24.13 | 0.613 |
| ViT-CNN | 14.01 | 20.78 | 0.453 | 16.98 | 17.08 | 0.409 |
| ViT-only | 7.75 | 22.98 | 0.541 | 17.23 | 20.53 | 0.511 |
| VRF-Net | **4.44** | **28.52** | **0.771** | **6.28** | **26.91** | **0.701** |

#### 4.3.2. On different pooling procedures

Moreover, we investigated how different pooling strategies affect system matrix recovery under various stride settings with our model. Specifically, we compared three approaches: using max pooling alone, using average pooling alone, and using the mixed strategy (max followed by average pooling) that served as the baseline in our main study. This is performed on the system matrix slice of the Open MPI Data acquired at a frequency of 380 kHz at a 4× scale factor.

| Pooling | Stride = 2 | Stride = 4 | nRMSE (stride =2, stride = 4) | pSNR (dB) (stride =2, stride = 4) | SSIM (stride =2, stride = 4) |
|---|---|---|---|---|---|
| **max** | 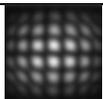 | 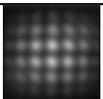 | (3.972, 6.421) | (30.43, 17.74) | (0.681, 0.432) |
| **average** | 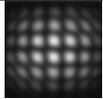 | 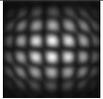 | (3.632, 4.871) | (32.56, 25.81) | (0.701, 0.678) |
| **Mixed (ours)** | 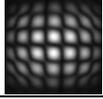 | 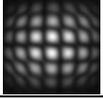 | (3.404, 4.812) | (38.06, 31.76) | (0.729, 0.671) |

**Fig. 12.** Comparison of recovered system matrix slices under different pooling strategies (max, average, and mixed) with stride = 2 and stride = 4.

As shown in Fig. 12, max pooling preserves sharp details but also amplifies noise, leading to higher error (nRMSE = 3.972, 6.421) and lower fidelity (pSNR = 30.43, 17.74; SSIM = 0.681, 0.432). Average pooling improves stability and similarity (SSIM = 0.701, 0.678) but oversmooths the image, reducing fine details. Mixed pooling, in contrast, introduces the strongest deterioration of fine frequency details, as reflected in its more aggressive downsampling. However, despite this greater frequency loss, our model successfully compensates for it, yielding the best overall recovery with the lowest errors (NRMSE = 3.404, 4.812), the highest pSNR (38.06, 31.76), and strong SSIM (0.729, 0.671). This demonstrates that while max or average pooling alone can be used, our framework is most robust when challenged by stronger fine-frequency deterioration, showing its ability to restore system matrix quality even under stronger downsampling.

### 4.3.3. On different model parameters

More ablation studies were performed to evaluate how the transformer model's parameters affect the recovered system matrix. By keeping the embedded dimension ($\varphi$) and MLP dimension constant, we explored different transformer configurations using the Open MPI data at a 2× resolution scale factor. When varying the number of heads, the layer depth was fixed at 3 (as in the basic study), and when varying layer depth, the number of heads was fixed at 8. To mitigate overfitting, a slightly lower learning rate ($10^{-8}$) was used when increasing heads, and a dropout of 0.1 was applied when increasing depth.

As shown in Table 9, increasing the number of heads clearly improved all metrics: 2 heads gave a high nRMSE of 9.042 and low SSIM of 0.501, 6 heads reduced nRMSE to 1.437 with SSIM 0.698, and our 8-head configuration achieved the best results (nRMSE 0.403, PSNR 39.08, SSIM 0.835). Similarly, deeper transformer layers generally enhanced performance: stacking 2 layers gave moderate results (nRMSE 2.678, SSIM 0.679), while 6 layers slightly improved nRMSE (0.401) and SSIM (0.850) but incurred higher computational cost. Our final model, using 3 layers and 8 heads, effectively balances reconstruction accuracy and efficiency, achieving metrics close to the deepest configuration while remaining practical for our model's aim.

**Table 9.** Ablation study exploring the effect of transformer attention heads and layer depth on the system matrix recovery performance (nRMSE, pSNR, SSIM) using Open MPI data at 2× resolution.

| Parameter | value | nRMSE | pSNR | SSIM |
|---|---|---|---|---|
| **head** | 2 | 9.042 | 19.12 | 0.501 |
| | 4 | 4.760 | 24.87 | 0.687 |
| | 6 | 1.437 | 29.60 | 0.698 |
| (ours) | **8** | **0.403** | **39.08** | **0.835** |
| **Layer depth** | Stacked 2 times | 2.678 | 29.64 | 0.679 |
| | Stacked 4 times | 0.583 | 29.97 | 0.703 |
| | Stacked 6 times | **0.401** | 37.04 | **0.850** |
| (ours) | **Stacked 3 times** | **0.403** | **39.08** | **0.835** |

### 4.4. Inference results

Furthermore, the pre-trained VRF-Net was utilized to analyze other simulated datasets, other than those originally employed for system matrix recovery. These datasets were constructed by extracting from simulated matrices that did not include coil sensitivity data. To further enhance the diversity of the data, random adjustments were made, such as adding noise, flipping, and sine wave distortion [42].

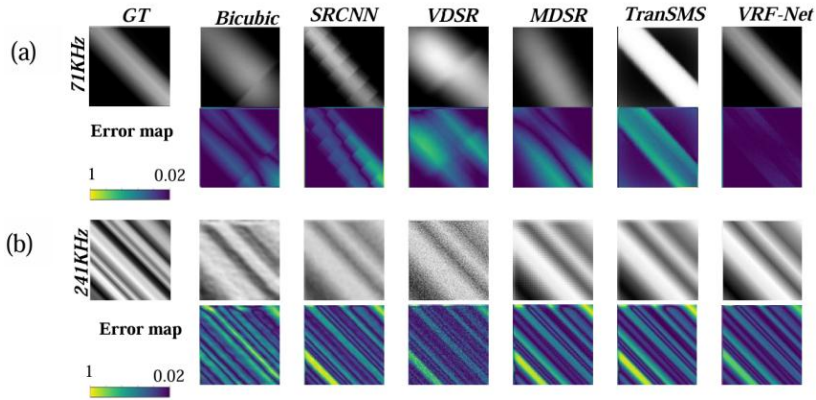

**Fig. 13.** The inference results for system matrix recovery for a scale factor of 2× as in (a) and 4× as in (b).

At the 2× scale (Fig. 13a), all models attempt to reconstruct the underlying structure, but clear differences emerge in terms of fidelity. Bicubic interpolation produces a smooth but overly blurred result, lacking fine detail. Classical CNN-based models such as SRCNN and VDSR introduce high-frequency noise or artifacts due to limited receptive field and shallow modeling. While MDSR improves structural recovery, some detail is still lost. TranSMS offers better visual sharpness and structure alignment but exhibits slight overshooting in brighter regions. In contrast, VRF-Net most closely matches the ground truth, maintaining clean boundaries and subtle variations. The accompanying error map highlights its minimal deviation, with residuals concentrated at high-frequency edges.

At the 4× scale (Fig. 13 b), the degradation is more severe, and most baseline methods struggle significantly. VDSR and SRCNN outputs become more distorted and noisier, while Bicubic produces overly smoothed representations. MDSR loses fine structure under stronger compression, and TranSMS—although more resilient—shows slight artifacts in detail reconstruction. VRF-Net remains stable, capturing both the broad structure and localized features. Its error map shows the lowest energy among all models, indicating effective suppression of misleading details and superior alignment with the ground truth.

Overall, these results confirm that VRF-Net generalizes well to unseen datasets and variable conditions. Its hybrid architecture, leveraging global attention and local residual enhancement, proves effective at identifying and reconstructing meaningful features even under distortion.

**Table 10.** The estimated average time (in seconds) for system matrix recovery for each of the competing methods with VRF-Net on the simulated dataset.

| *Method* | *2×* | *4×* | *8×* |
|---|---|---|---|
| **Bicubic** | 1.50 | 2.09 | 2.07 |
| **SRCNN** | 0.65 | 0.76 | 0.78 |
| **VDSR** | 26.78 | 27.8 | 27.83 |
| **MDSR** | 78.78 | 79 | 82.18 |
| **TranSMS** | 170.18 | 172.7 | 176.87 |
| **VRF-Net** | 7.58 | 9.09 | 11.69 |

From Table 10, the traditional methods like Bicubic interpolation are the fastest, taking just around 1.5 to 2 seconds. Similarly, SRCNN, a shallow CNN model, performs slightly faster at under one second. However, while both are fast, they offer limited accuracy, especially when it comes to recovering complex structures in the system matrix.

On the other end of the spectrum, more advanced models like VDSR and MDSR show significantly higher inference times, approximately 27 seconds for VDSR and up to 82 seconds for MDSR, due to their deeper residual networks and multi-scale designs. The TranSMS model, which incorporates heavy transformer components, is the slowest, taking up to 177 seconds. In contrast, the proposed VRF-Net demonstrates a much more practical balance. It completes inference in just 7.58 seconds at 2× and 11.69 seconds at 8×, making it far more efficient than the deeper models while maintaining high reconstruction quality.

## 5. Discussion

In this study, we proposed VRF-Net, a hybrid deep learning framework that integrates vision transformers for global context modeling with residual feature blocks for local detail refinement, specifically designed to super-resolve MPI system matrices. Our results provide strong support for the initial hypothesis; the quantitative metrics and qualitative reconstructions confirm that VRF-Net consistently outperformed baseline models in terms of pSNR, SSIM, and nRMSE, with clear improvements in resolving structural details of the resolution phantom. Furthermore, the t-SNE analyses show well-separated feature clusters across training and testing, indicating that the hybrid model effectively learns both global and local information as hypothesized. Finally, the results across both the experimental Open MPI dataset and the simulated dataset with variable coil sensitivities validate our second hypothesis of cross-dataset generalization.

The model was trained using a paired image super-resolution approach on both the Open MPI dataset and a custom-simulated dataset. Special care was taken to tune the model to balance two competing goals: reducing noise introduced by dual downsampling and preserving subtle yet essential features that are critical for accurate system matrix recovery [43]. The use of dual-stage downsampling was intentionally designed to test the model's robustness; while it improves resilience to degradation, it also places a greater demand on the feature extraction mechanism [44]. While the pooling operations are effective for reducing dimensionality and highlighting dominant features, they can lead to over-smoothing, which may limit the model's ability to recover high-frequency details, especially under complex or noisy conditions [45].

The first stage of downsampling mimics resolution loss due to hardware or sampling constraints, whereas the second stage introduces additional signal distortion that behaves like random noise. To further approximate real-world acquisition conditions, Gaussian perturbations with a standard deviation of 0.01–0.05 (relative to normalized matrix values) were introduced during inference, along with sinusoidal distortions to emulate field instabilities.

The "noise" and "artifacts" that VRF-Net learns to correct are not abstract but are deliberately modeled to reflect real MPI conditions. To capture stochastic noise, the Open MPI dataset was restricted to system matrices with SNR ≥ 10 for stable training, while dual-stage downsampling was applied to deliberately suppress 25–50% of high-frequency components. However, the use of SNR ≥ 10 for the Open dataset also introduces a limitation: excluding lower-SNR samples may make the model less effective in the in-vivo conditions, where signal degradation is common. Future versions of the model could benefit from incorporating a broader SNR spectrum to improve generalization to noisy clinical environments.

Structured artifacts were modeled by simulating coil sensitivity variations (0.5–1 mV/μT) using Biot–Savart–based profiles, which produce direction-dependent distortions in the system matrix. These sensitivity-induced variations alter spatial encoding and create frequency-dependent biases similar to those observed in practice. Together, these degradations mean that VRF-Net is not only "enhancing" resolution by recovering suppressed high-frequency details but also "denoising" by reducing noise-like fluctuations and correcting coil-induced distortions. In effect, the model learns to deliver sharper, artifact-free system matrices that preserve fine spatial structures while remaining robust against common MPI noise sources. This two-step degradation strategy encourages the model to develop robust feature recovery under imperfect conditions, effectively training it to generalize beyond the clean data it was initially exposed to.

While it doesn't replace clinical testing, it provides a structured way to inject noise resilience into the learning process without compromising stability[46].

One key purpose of dual downsampling is to simulate realistic degradation and compensate for using high-SNR ($\geq 10$) training data to enhance the model's robustness in noisy environments while maintaining controlled training conditions. Since we restricted the training set for the Open dataset to system matrices with relatively clean signals (SNR $\geq 10$), it was essential to introduce a mechanism that forces the model to also learn from more challenging, noise-like scenarios without directly using unstable low-SNR data[47]. The dual-stage downsampling process primarily degrades the high-frequency information rather than eliminating it. Max pooling emphasizes dominant local responses, while average pooling smooths variations, together attenuating fine details in a way that mimics realistic scanner-induced degradation. Although some irretrievable information loss occurs, much of the high-frequency content remains in degraded form.

VRF-Net is trained to exploit these degraded signals and recover them into a high-resolution system matrix. Although direct 3D processing of Open MPI data using the dimensional decoupling could theoretically model cross-slice dependencies more explicitly, our results demonstrate that the 2D-slice extracted approach achieves high structural fidelity (SSIM up to 0.96) without noticeable discontinuities in the recovered system matrix. Future work may integrate 3D attention mechanisms or hybrid slice-stacking strategies to further enhance spatial coherence across slices.

The architectural complexity of VRF-Net lies in its attempt to balance global structural understanding and fine-detail recovery. This is achieved by integrating a Vision Transformer (ViT) with a series of 10 residual feature blocks. These residual blocks enhance the model's capacity to restore fine-grained details, even from heavily downsampled or noisy data. However, this deep integration also requires careful calibration; too many residual layers can increase computational cost, and misalignment between the residual and transformer features may limit effectiveness. Achieving harmony between local feature refinement (via residual blocks) and global attention modeling (via the ViT) is therefore critical for optimal performance.

The transformer module provides high model capacity by capturing long-range dependencies and global frequency features, while the residual feature network constrains learning to localized residual corrections, effectively regularizing the model. This design reduces the likelihood of overfitting compared with a pure high-capacity transformer.

To further mitigate overfitting, we applied SNR-based filtering (SNR $\geq 10$) to the Open MPI dataset and extensive data augmentation (flipping, Gaussian perturbations, sinusoidal distortions) to both datasets. These choices limit the model's ability to memorize training data and encourage generalization. The observed convergence of training and validation losses, together with consistently high SSIM (0.96) on the reconstructed images, as in Table 6, suggests that overfitting was not a dominant issue in our experiments. A practical concern with vision transformers is the quadratic scaling of self-attention with input size, which can lead to high computational and memory costs, especially for 3D system matrices or large-scale MPI acquisitions. In this work, we addressed this by feeding VRF-Net with 2D slices rather than full 3D matrices, significantly reducing input sequence length and computational burden while still capturing spatial correlations. Furthermore, we employed patch embedding to reduce the effective sequence length before self-attention and designed VRF-Net as a hybrid model, where convolutional residual feature blocks handle local refinement and the transformer captures global dependencies. This division of labor reduces the demand on the attention mechanism and ensures stable GPU memory usage during training and inference.

Looking ahead, scaling VRF-Net to larger 3D system matrices can be managed by extending these design choices with window-based or hierarchical attention mechanisms to constrain the receptive field, as well as mixed-precision training and optimized GPU memory handling. These strategies make it feasible to adapt the framework to more computationally intensive MPI datasets.

Regarding error behavior under varying SNR conditions, MPI signals degrade approximately in proportion to 1/SNR, where lower SNR introduces larger variance in high-frequency components. In our framework, this manifests as a broader spread of reconstruction error. Specifically, empirical results showed that SSIM decreased by ~5–7% and nRMSE increased by ~10–12% when Gaussian noise was added at 0.01–0.05 relative variance. These trends are consistent with theoretical expectations that error bounds widen as SNR decreases, since VRF-Net must increasingly rely on learned priors rather than signal content. While VRF-Net demonstrated robustness within the tested SNR range, we note that extending training to cover lower-SNR system matrices for the real-world scenarios remains an important direction for future work.

In MPI, the time derivative of magnetization, which reflects how magnetic nanoparticles respond to dynamic magnetic fields, is directly influenced by coil sensitivity. Coil sensitivity determines how strongly the receiver detects magnetization changes in different spatial directions, and it varies across the field of view and frequency spectrum. Simulated data plays an important role in this study by allowing precise control over these coil sensitivity profiles. As the system matrix is fundamentally shaped by coil responses, simulating this behavior provides a meaningful and customizable training environment [24].

The coil sensitivity profiles $S_x$, $S_y$ and $S_z$ were simulated using the Biot-Savart law (as detailed in Algorithm 1 and Eq. 8), which models the magnetic field generated by an ideal circular current loop. This method intentionally introduces spatial variations in sensitivity that are a function of the coil geometry and the voxel's position relative to the coil, as shown in Figure 5. These simulated variations are crucial for training the model to learn the fundamental coupling between spatial encoding and coil performance. However, it is important to note that this model operates under idealized assumptions (e.g., perfect circular geometry, uniform current, and no electromagnetic interference or hardware imperfections). Real-world coils exhibit more complex non-uniformities, edge effects, and mutual coupling not captured here. While this simplification allows for a controlled and reproducible investigation into the VRF-Net's ability to handle sensitivity-induced variations

Looking ahead, we plan to extend our simulation to include more realistic coil models, accounting for non-idealities such as hardware misalignment, non-uniform winding, or thermal noise factors that are commonly observed in real MPI systems. Additionally, we intend to explore multiscale feature learning architectures, such as feature pyramids or hierarchical networks. These techniques enable the model to learn both coarse and fine spatial patterns across different resolutions, thereby improving its ability to recover high-resolution system matrices while maintaining large-scale consistency.

While our results confirm the effectiveness of VRF-Net for super-resolving MPI system matrices, some limitations should be acknowledged. First, the absence of in vivo validation introduces additional variability, such as heterogeneous tissue properties and non-uniform magnetic susceptibility, which were not accounted for in this simulation-based study [48]. To mitigate this gap, several strategies can be pursued in future work, such as domain adaptation and transfer learning approaches can be applied, where a model trained on simulated and phantom data is fine-tuned using smaller amounts of experimental or preclinical in vivo data. Second, the Open MPI dataset provides an excellent controlled environment for initial testing. However, real-world medical imaging is more complex. Factors such as the variable magnetic properties of human tissues, patient motion, and more complicated nanoparticle behavior could affect the model's performance in clinical settings. These factors can influence both the signal response and the structure of the system matrix, potentially impacting the model's clinical reliability. Third, while our simulated dataset was generated using variations in magnetic nanoparticle size, FOV, and coil sensitivity, only the coil sensitivity variations were used in the training of the current model. Incorporating more complex and realistic physical parameters will be essential for advancing toward clinically robust MPI reconstruction.

## 6. Conclusion

We propose an MPI system matrix recovery approach employing a hybrid deep learning framework that combines the global attention capabilities of vision transformers with the local refinement power of residual feature blocks to recover high-resolution MPI system matrices. Our approach demonstrated clear advantages over existing state-of-the-art techniques: the recovered system matrices consistently produced higher pSNR and SSIM values with lower nRMSE, and qualitative reconstructions of the resolution phantom showed sharper structural details and reduced noise, as shown in Table 11. These findings highlight the effectiveness of VRF-Net in restoring fine frequency components that are often lost in degraded system matrices, ultimately leading to more reliable MPI image reconstruction.

Looking forward, several future research directions remain open. First, the model should be validated on a broader range of experimental MPI datasets beyond the currently available Open MPI data, including data acquired under different scanner settings and phantoms. Second, extending the framework to in vivo data will be an important step toward clinical translation. Third, exploring real-time implementations and optimizing the computational efficiency of VRF-Net will make it more practical for preclinical and clinical MPI applications.

Overall, VRF-Net provides not only a methodological advancement in MPI system matrix recovery but also a foundation for future studies aiming to bridge the gap between simulation-based research and real-world clinical imaging.

**Table 11.** Quantitative comparison of the proposed VRF-Net with state-of-the-art methods for MPI system matrix recovery (SMR) and image reconstruction (IR). The table summarizes pSNR, nRMSE, and SSIM values reported in the literature, along with their corresponding methodological approaches. For each study, the best reported performance under comparable conditions is listed to highlight differences in accuracy, reconstruction quality, and recovery strategies.

| Study | pSNR (dB) | nRMSE | SSIM | Comments |
|---|---|---|---|---|
| **CCS** [21] | 22.12 (Shape phantom, δ=0.2) | - | 0.97(Shape phantom, δ=0.2) | Coded Scenes with different filling rates (IR Only) |
| **3D-SMRnet**[19] | 55.03 (8×), (shape phantom) | 0.018 (8×), (shape phantom) | 0.99 (8×), (shape phantom) | 3D SM recovery (IR Only) |
| **LIIF**[18] | - | 12.77% | - | Implicit Function (SMR Only) |
| **TranSMS**[17] | 38.54 (2×)/ 24.89(2×) _ | 3.15% (2×)/ 6.03% | - | Transformer (SMR/IR) |
| **DIP-SM**[23] | 34.6 (2×)/ 38.55 (2×) | 0.115(2×) /0.083 (2×) | 0.975 (2×)/0.993 | Deep image prior (SMR/IR) |
| **ProTSM**[16] | ~35.90 (2×)/41.43(2×) | ~3.08% (2×)/0.86% (2×) | 0.737 (Only for IR) | Progressive Pretraining (SMR/IR) |
| **DEQ-MPI**[26] | 37.7±3.2 (varies by SNR) | - | 86.3% ±5.5 | Deep equilibrium model (IR Only) |
| **MPIGAN**[25] | 22.57 | 0.0173 (RMSE) | 0.968 | End-to-End GAN (IR Only) |
| **RETNet**[24] | 31.72 (noisy condition) | 0.012 (noisy condition) | 0.940 (noisy condition) | Transformer (X-Space) (IR Only) |
| **DERnet** [49] | 28.9 ± 3.31 (30dB) | 0.146 ± 0.0548 (30dB) | 0.984 ± 0.0114 (30dB) | End-to-End Network (IR) |
| **Nonconvex Regularization** [50] | 57.8 (Stenosis Phantom) | 0.001 (Stenosis Phantom) | 0.99 (vessel Phantom) | Nonconvex ADMM (IR Only) |
| **3D-ISSRnet** [51] | 35.33 (2×) | 0.017 (2×), (RMSE) | 97 % (2×), (Dice) | Iterative up-and-down sampling super-resolution (IR Only) |
| **ZeroShot-ℓ1-PnP** [52] | 37.54 concentration phantom) | - | 0.954 (Shape phantom) | Plug-and-Play (IR Only) |
| **Multi-task TransGAN**[27] | 39.37 | 0.0018 | 0.994 | Multi-task GAN (IR Only) |
| **INR for Arbitrary Scale SR** [53] | 39.67(2×) (Only for IR, Concentration phantom) | 0.033 (2×) (Only for SMR) | 0.939 (3×), (Only for IR, Resolution Phantom) | Continuous implicit neural representation |
| **Ours** | 39.08 (2×)/41.58 (resolution phantom) | 0.403 (2×)/1.79 (2×) (resolution phantom) | 0.835 (2×)/0.960 | ViT + Residual Feature Network (SMR/IR) |

**Ethics Statement**

All procedures were performed in compliance with relevant laws and institutional guidelines. This study did not involve human participants, animals, or identifiable personal data. Therefore, ethical approval and informed consent were not required.

**Acknowledgments**

This work is supported in part by the Key R&D Program of Shandong Province, China 2022CXGC010501, and the National Natural Science Foundation of China under Grant No. 82227802.